\documentclass[11pt]{article}

\usepackage{amsfonts}
\usepackage{amsmath}
\usepackage{amssymb}
\usepackage{amsthm}
\usepackage{mathrsfs}
\usepackage{graphicx}

\usepackage[]{algorithm2e}

\theoremstyle{plain}

\theoremstyle{definition}

\numberwithin{exercise}{section}
\numberwithin{equation}{section}
\numberwithin{theorem}{section}
\numberwithin{problem}{section}

\numberwithin{figure}{section}

\DeclareMathOperator{\Int}{int}

\newcommand{\bs}[1]{{\boldsymbol{#1}}}

\newcommand{\R}{\mathbf{R}}

\newcommand{\D}{\,\mathrm{d}}

\newcommand{\IP}[2]{\left\langle#1\,,#2\right\rangle}

\begin{document}

\title{Food webs and the principle of evolutionary adaptation}

\author{Alexander S. Bratus$^{1,2,}$\footnote{e-mail: alexander.bratus@yandex.ru}$\,\,$, Anastasiia V. Korushkina$^{3}, $ Artem S. Novozhilov$^{4},$\footnote{e-mail: artem.novozhilov@ndus.edu} \\[3mm]
\textit{\normalsize $^\textrm{\emph{1}}$Russian University of Transport, Moscow 127994, Russia}\\[0mm]
\textit{\normalsize $^\textrm{\emph{2}}$Moscow Center of  Fundamental and  Applied Mathematics,}\\[-1mm]
\textit{\normalsize Lomonosov Moscow State University, Moscow 119992, Russia}\\[0mm]
\textit{\normalsize $^\textrm{\emph{3}}$Faculty of Computational Mathematics and Cybernetics,}\\[-1mm]
\textit{\normalsize Lomonosov Moscow State University, Moscow 119992, Russia}\\[0mm]
\textit{\normalsize $^\textrm{\emph{4}}$Department of Mathematics, North Dakota State University, Fargo, ND, 58108, USA}}

\date{}

\maketitle

\begin{abstract}A principle of evolutionary adaptation is applied to the Lotka--Volterra models, in particular to the food webs. We present a relatively simple computational algorithm of optimization with respect to a given  criterion. This algorithm boils down to a sequence of easy to solve linear programming problems. As a criterion for the optimization we use the total weighted population size of the given community and an ecological fitness, which is an analogue of the potential energy in physics. We show by computational experiments that it is almost always possible to substantially increase the total weighed population size for an especially simple food web --- food chain; we also show that food chains are evolutionary unstable under the given optimization criteria and, if allowed, evolve into more complicated structures of food webs.

\paragraph{\small Keywords:} Lotka--Volterra model, food chain, food web, fitness maximization

\paragraph{\small AMS Subject Classification: } 92D15, 92D25, 92D40

\end{abstract}

\section{Controlling food chains}
The replicator equations and Lotka--Volterra models of ecological communities are among the most basic and classical mathematical models in mathematical biology, e.g., \cite{hofbauer1998ega}. The methods of physics and statistical physics in particular have a long history being applied to these models, e.g. \cite{goel1971volterra}. In the vast majority of cases, however, it is assumed that the parameters in these systems of ordinary differential equations are constant for each particular system realization. That is, the evolution of distributions in the replicator equation and the evolution of the population sizes in the Lotka--Volterra equations is usually conditioned on the premise that the given set of parameters is fixed for given circumstances. Clearly, the conditions may change, and hence the evolution of the quantities we are interested in also changes. However, such case is usually considered as a different possible realization of the same system.

On the other hand one of the most basic and foundational principles in physics is that the observed regularities are the consequence of some extremal principle (e.g., least action principle). For many years there was a search for a similar principle(s) to explain the observed regularities in ecological communities (see, e.g., a historical discussion in \cite{svirezhev1978stability}). Another possible approach to this problem is first to formulate a mathematical model, postulate an extremal principle (define a quantity to be maximized or minimized) and then study the regularities which will form under this principle. In the present text we follow this rout.

In \cite{bratus2018evolution} it was suggested to look at the replicator equation from an evolutionary prospective, allowing for the changes of the parameters within the same system realization; the basic assumption was the separation of the time scales whereas during the fast time dynamics the system settles at an asymptotic regime, and during the slow time dynamics the system is allowed to change its parameter values according to some prescribed evolutionary principle (usually the fitness maximization). This approach was further extended and illustrated by other examples of replicator equations in \cite{drozhzhin2021fitness} and \cite{samokhin2020open}. The Lotka--Volterra equations, being a close relative to the replicator equation, are equally well suited for this approach of evolutionary adaptation, and in the present text we show how to implement it using a specific example of the Lotka--Volterra equations ---  the so-called \textit{food chains} and \textit{food webs} (e.g., \cite{cohen2012community}).

Our starting point is the dynamical system that models \textit{food chain}, i.e., we assume that there is a basal species at the bottom of the food chain, there is a predator for this basal species, which is the only prey for another predator, and so on. Taking the intra-specific competition into account we end up with the system of ordinary differential equations
\begin{equation}\label{eq1:1}
   \dot u_i=u_i\Bigl(\rho_i-\bigl(\bs{Au}\bigr)_i\Bigr),\quad i=1,\ldots,n,
\end{equation}
where $\rho_1=r_1>0, \rho_j=-r_j, r_j>0,\,j=2,\ldots,n,$ are the Malthusian parameters, in particular $\rho_2,\ldots, \rho_n$ are the death rates of the predators in the chain; matrix $\bs A$ has the form
\begin{equation}\label{eq1:2}
    \bs A=\begin{bmatrix}
            a_{11} & a_{12} & 0 & \ldots & 0 & 0 \\
            -a_{21} & a_{22} & a_{23} & \ldots & 0 & 0 \\
            \vdots &  &  &  &  &\vdots  \\
            0 & 0 &\ldots &0& -a_{n,n-1} & a_{nn} \\
          \end{bmatrix},
\end{equation}
where all the parameters are positive; finally, $(\bs u)_i$ is the $i$-th element of the vector $\bs u=(u_1,\ldots, u_n)$.

It is well known \cite{harrison1979global,so1979note,hofbauer1998ega} that if there exists an equilibrium $\bs{\hat u}$ of \eqref{eq1:1} in the interior of $\R^n_+$ then this equilibrium is globally stable in the sense that any orbit starting in the interior of $\R^n_+$ converges to $\bs{\hat u}$.

First we consider the problem to control a given food chain through the choice of the death rates $r_2,\ldots,r_n$. From the practical point of view this is a natural thing to look into, since in this case we can imagine that we assign quotes to harvest or hunt on particular species, which directly leads to changes in the death rates. If we, say, increase $r_j$, it means that we increase the quotes to harvest the $j$-th species; similarly, if we decrease $r_j$, it implies that we employ some strategies to boost the $j$-th species reproduction.

Within given above interpretation it is a reasonable goal to maximize the (weighted) total population size at a fixed time moment $t=T$:
\begin{equation}\label{eq1:3}
    S_p(\bs u(t))=\sum_{i=1}^n p_i u_i(t)\longrightarrow \max_{t=T},
\end{equation}
where $p_i\geq 0$ are given weights.

To the best of our knowledge, this problem was analyzed mathematically for the first time in \cite{svirezhev1972}. In this work the functional $S_p(\bs u)$ was maximized for the predator--prey model for the given constant $T$ and given initial conditions, which poses some computational difficulties. In the present text we solve the same problem using an additional hypothesis of time separation in our model, similar to what was done for certain classes of replicator equations in \cite{bratus2018evolution,drozhzhin2021fitness,samokhin2020open}. This hypothesis allows a significant simplification in solving problem \eqref{eq1:3}.

To wit, we assume that there are two time scales in our problem: fast system dynamics when all the parameters are assumed constant, and slow dynamics, in which we allow changes in the death rates (more general case will be considered below). Mathematically, we have
\begin{equation}\label{eq1:4}
    \frac{\D u_i}{\D t}(t;\epsilon t)=u_i(t;\epsilon t)\Bigl(\rho_i(\epsilon t)-\bigl(\bs {Au}(t;\epsilon t)\bigr)_i\Bigr),\quad i=1,\ldots, n,
\end{equation}
where $\epsilon>0$ is a small parameter. Introducing $\tau=\epsilon t$ yields
\begin{equation}\label{eq1:5}
    \frac{\D u_i}{\D \tau}(\tau/\epsilon;\tau)=u_i(\tau/\epsilon;\tau)\Bigl(\rho_i(\tau)-\bigl(\bs {Au}(\tau/\epsilon;\tau)\bigr)_i\Bigr),\quad i=1,\ldots, n.
\end{equation}
For the finite values $0\leq t\leq T$, due to the smallness of $\epsilon$, the dynamics of \eqref{eq1:1} is close to the dynamics of \eqref{eq1:4}. Taking the limit $\epsilon\to 0$ implies
\begin{equation}\label{eq1:6}
    \hat u_i(\tau)\Bigl(\rho_i(\tau)-\bigl(\bs{A\hat u}(\tau)_i\bigl)\Bigr)=0,\quad i=1,\ldots,n,
\end{equation}
where $\hat{u}_i(\tau)=\lim_{\epsilon\to 0}u_i(\tau/\epsilon;\tau)$. If we impose the restriction that $\hat{u}_i\neq 0$ for any $i$, the process of evolutionary changes of the system boils down to consideration of the system of linear equations
\begin{equation}\label{eq1:7}
    \bs{A\hat u}(\tau)=\bs{\rho}(\tau),\quad \bs \rho(\tau)=\bigl(r_1(\tau),-r_2(\tau),\ldots,-r_n(\tau)\bigr),
\end{equation}
which depends on the slow time $\tau$. As a result, we obtain a related but different maximization problem
\begin{equation}\label{eq1:8}
    S_p(\bs{\hat u}(\tau))=\sum_{i=1}^n p_i\hat u_i(\tau)\longrightarrow\max_{\rho(\tau)},
\end{equation}
on all possible solutions to \eqref{eq1:7}, where vector $\bs \rho$ must satisfy certain constraints. Specifically, we assume that parameter $r_1$ is fixed, i.e., we do not allow harvesting the basal species. The rest of the parameters must satisfy
\begin{equation}\label{eq1:9}
    m_i\leq r_i(\tau)\leq M_i,\quad i=2,\ldots, n,
\end{equation}
where $m_i,M_i$ are given positive constants.

To further represent problem \eqref{eq1:7}--\eqref{eq1:9} in the form of evolutionary adaptation, we divide this problem into a sequence of steps, at each of which a linear programming problem should be solved. Strictly speaking this step is not necessary for the problem at hands, but will be essential for the more complicated problems, considered in the following.

Consider an allowable perturbation of vector $\bs \rho$ in the form
$$
\bs \rho+\Delta\bs \rho=\bigl(r_1,-(r_2+\Delta r_2),\ldots,-(r_n+\Delta r_n)\bigr),
$$
assuming
$$
\sum_{i=2}^n \Delta r_i\leq \varepsilon, \quad \varepsilon>0,
$$
where $\varepsilon$ is a sufficiently small given constant. Then the equilibrium vector $\bs{\hat u}$ will get a perturbation $\Delta \bs{\hat u}$. Clearly, the equation connecting $\Delta\bs \rho$ and $\Delta \bs{\hat u}$ is the linear system
\begin{equation}\label{eq1:10}
    \bs A\Delta\bs{\hat u}=\Delta\bs \rho.
\end{equation}
Consider an auxiliary system
\begin{equation}\label{eq1:11}
    \bs A^\top \bs{v}=\bs p.
\end{equation}

After taking the scalar product of both sides of \eqref{eq1:10} with $\bs v$ we obtain
$$
\IP{\Delta\bs \rho}{\bs v}=\IP{\bs A\Delta\bs{\hat{u}}}{\bs v}=\IP{\Delta\bs{\hat u}}{\bs A^\top\bs v}=\IP{\Delta\bs{\hat u}}{\bs p},
$$
from where
$$
\Delta S_p=\IP{\Delta \bs{\hat u}}{\bs p}=-\sum_{i=2}^nv_i\Delta r_i.
$$

Let $K^{+}$ ($K^{-}$) be the subsets of the set of indexes $\{2,\ldots,n\}$, for which the components of $\bs v$ that solves \eqref{eq1:11} are positive (negative) and, assuming we have no zero components, $K^{+}\bigcup K^{-}=\{2,\ldots,n\}$.

The form of the expression $\Delta S_p$ immediately implies that if both of the sets $K^{+}$ and $K^{-}$ are nonempty then there is a potential way to increase the value of $S_p$ by either choosing $\Delta r_i<0$ for $i\in K^{+}$ or $\Delta r_i>0$ for $i\in K^{-}$. Moreover, if $K^{+}=\emptyset$ ($K^{-}=\emptyset$) then increase of the total population size at the expense of increasing harvest of some specific species (decreasing it) is impossible.
We say ``a potential way to increase the value of $S_p$'' because the analysis of perturbations $\Delta r_i$ and $\Delta{\hat u}_i$ should also include the analysis of given constraints. In particular, recall that $M_i\geq r_i\geq m_i$ for all $i$. So, if we find that, for instance, $r_i\leq m_i$ then one must add that $\Delta r_i\geq 0$ with certain changes for the expression for $\Delta S_p$. Similarly, since our hypothesis of time separation is based on the fact that the internal equilibrium of the food change is globally stable, if it exists, we also require that all $\hat u_i\geq \delta>0$, and if for some index $i$ we already approached the coordinate plane, i.e., $\hat u_i=\delta$ then one must additionally require that
$$
\Delta \hat{u}_i=\Bigl(\bs{A}^{-1}\Delta\bs \rho\Bigr)_{i}\geq 0.
$$

Taking everything together, we obtain an efficient numerical procedure to solve problem \eqref{eq1:8}, and hence, with high degree of accuracy, approximate significantly more computationally involved problem \eqref{eq1:3}.

Here is the basic algorithm. We start with fixing the values of $m_i,M_i,i=2,\ldots,n,\varepsilon>0,\delta>0$. Let $\bs \rho^0$ satisfies the given constraints. We first solve the auxiliary problem \eqref{eq1:11}, find $\bs v$, and consider the linear programming problem
\begin{equation}\label{eq1:12}
    \Delta S_p=-\sum_{i=2}^n v_i\Delta r_i^0,\quad \sum_{i=2}^n \Delta r_i^0\leq \varepsilon, \quad i=2,\ldots, n,
\end{equation}
and such that the vectors $\bs \rho^0+\Delta\bs \rho^0$ and $\bs{\hat u}$ are allowable. It means that $\bs \rho^0+\Delta\bs \rho^0$ within the given constants $m_i,M_i$, and $\bs{\hat{u}}^0+\Delta \bs{\hat{u}}^0$ is away from the boundary of $\R^n_+$ at least $\delta$ units. If we are capable to find such $\Delta\bs\rho^0$ we compute
$$
\bs \rho^1=\bs \rho^0+\Delta \bs \rho^0
$$
and repeat the previous step, until we can make an allowable change. This problem is convex because of the convexity of the set of imposed constraints and linearity of problem \eqref{eq1:7}, and therefore its solution always exists, although it could be non unique.

Here is an example. We note that we use this example only for illustrative purposes to show how our algorithm of evolutionary adaptation works, without pretending that a real food chain is considered. We consider the case of $n=10$ species in the food chain.

Let
\begin{equation}\label{eq:A}
\bs A=
\left[
\begin{array}{cccccccccc}
 1 & 4 & 0 & 0 & 0 & 0 & 0 & 0 & 0 & 0 \\
 -4 & 1 & 6 & 0 & 0 & 0 & 0 & 0 & 0 & 0 \\
 0 & -7 & 1 & 5 & 0 & 0 & 0 & 0 & 0 & 0 \\
 0 & 0 & -6 & 1 & 3 & 0 & 0 & 0 & 0 & 0 \\
 0 & 0 & 0 & -8 & 1 & 6 & 0 & 0 & 0 & 0 \\
 0 & 0 & 0 & 0 & -7 & 1 & 9 & 0 & 0 & 0 \\
 0 & 0 & 0 & 0 & 0 & -2 & 1 & 3 & 0 & 0 \\
 0 & 0 & 0 & 0 & 0 & 0 & -4 & 1 & 5 & 0 \\
 0 & 0 & 0 & 0 & 0 & 0 & 0 & -2 & 1 & 9 \\
 0 & 0 & 0 & 0 & 0 & 0 & 0 & 0 & -3 & 1 \\
\end{array}
\right]
\end{equation}
and
\begin{equation}\label{eq:B}
\bs\rho=(16, -1, -1, -1, -1, -1, -1, -1, -1, -2)
\end{equation}
in the problem $\eqref{eq1:1}$. We choose the following weights
\begin{equation}\label{eq:C}
\bs p=(10, 20, 30, 40, 50, 60, 10, 10, 10, 10).
\end{equation}

For the given parameter values we find, keeping one digit after the decimal point, that
$$
\bs{\hat{u}}=(4.2, 2.9, 2.2, 3.5,  2.8, 4.0, 1.6, 1.8, 0.7, 0.2)\in \R^n_+
$$
solves \eqref{eq1:10}, and, solving \eqref{eq1:11},
$$
\bs v=(48.1,   9.5,  26.0,   8.9,  12.4,  -1.6,    6.3,  -4.5,   2.2, -10.0).
$$
Therefore it looks like there is a potential to improve the value of $S_p$, which is initially $S_p(0)=731$.

The constraints for the problem are
$$
0.1\leq r_i\leq 3,\quad i=2,\ldots,n,
$$
$$
\hat{u}_i>\delta=0.001,\quad i=1,\ldots,n,
$$
and, finally, $\varepsilon=0.01.$

\begin{figure}[!t]
\begin{center}
\includegraphics[width=0.5\textwidth]{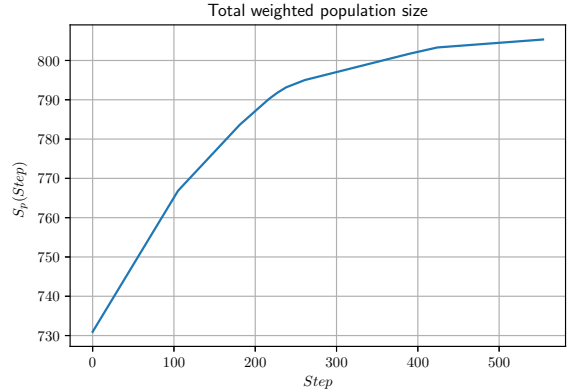}
\end{center}
\caption{The change in the value of functional $S_p$ given by \eqref{eq1:8} at each step of evolutionary algorithm in Section 1.}\label{fig:1}
\end{figure}

\begin{figure}[!b]
\begin{center}
\includegraphics[width=0.5\textwidth]{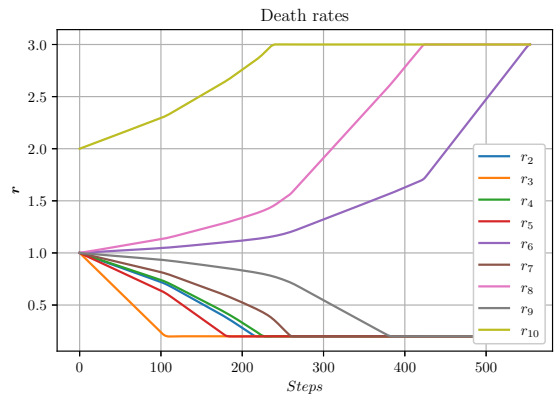}
\end{center}
\caption{The change in the death rates $\bs r$ at each step of evolutionary algorithm in Section 1.}\label{fig:2}
\end{figure}

Now we can run a sequence of steps, at each of which a linear programming problem is solved, until our solution moves outside of the allowable set. For the given set of parameters the algorithm runs for approximately 550 steps, at which the value of $S_p$ is 805, see Fig. \ref{fig:1}.
As we see in the coordinates of vector $\bs v$ for the given example, the algorithm tries to increase the sixth, eighth, and tenth death rates and decrease all others, see Fig. \ref{fig:2}.

In general what this simple numerical experiment shows is that if one is capable to modulate the death rates of a given food chain within a reasonable interval of allowable values, it is almost always possible (and was \textit{always} possible in our numerical experiments) to increase the weighted total population size at the equilibrium. Moreover, an extremely simple numerical procedure of finding the auxiliary vector $\bs v$ in \eqref{eq1:11} indicated which coefficients of the vector $\bs\rho$ in \eqref{eq1:1} should be increased and which decreased during the evolutionary steps.

As a final remark here, we note that from a practical point of view it it the harvesting of the top predator in a given food chain that of the most interest. This specific situation can be easily modeled with the described approach by setting vector $\bs p$ to $(0,\ldots,0,1)$.

\section{From food chains to food webs}

In the previous section we analyzed only food chains, i.e., communities, for which the interaction graph of matrix $\bs A$ has the simplest linear structure, where any node, except for the first and the last ones, has only two neighbors, one is a prey and another one is a predator. In reality these graphs are certainly more complicated and represent what is generally called \textit{wood web}. Let us ask therefore the following question: Is it advantageous for a given food chain to modify itself (i.e., to add new trophic interactions) under given circumstances (evolutionary constraints)?

Mathematically it means that now we allow evolutionary changes in the elements of matrix $\bs A$. For simplicity, we first assume that the predator with number $k$ has the ability to feed on any species with numbers $k-j,j=2,\ldots,k-1$, where the basal species has the number $1$. For this computation we fix vector $\bs \rho$, and the elements of matrix $\bs A$ are now chosen from the allowable set $\sum_{i,j=1}^n a_{ij}^2\leq K$ for some prescribed constant $K>0$. In this case the elements of the perturbation $\bs A+\Delta\bs A$ of the original matrix $\Delta A$ must be negative, if they are under the main diagonal, and positive if they are above the main diagonal. As before, to use the principle of the time scale separation, we assume that each perturbation $\Delta a_{ij}$ satisfies
\begin{equation}\label{eq2:1}
|\Delta a_{ij}|\leq \varepsilon
\end{equation} for some given constant $\varepsilon$.

We also call any perturbation $\Delta\bs A$ from the set \eqref{eq2:1} allowable, if the spherical norm of matrix $\bs A$ is non-increasing, which, up to the second order, means that
\begin{equation}\label{eq2:2}
    \sum_{i,j=1}^n a_{ij}\Delta a_{ij}\leq 0.
\end{equation}
Assuming that all the perturbations are proportional to $\varepsilon>0$ we have, up to the order $O(\epsilon^2)$,
\begin{equation}\label{eq2:3}
    \bs A\Delta\bs{\hat u}+\Delta \bs A\bs{\hat u}=0.
\end{equation}
Multiplying the equality \eqref{eq2:3} by $\bs v$ that solves \eqref{eq1:11}, we find that
\begin{equation}\label{eq2:4}
    \Delta S_p=\IP{\Delta\bs{\hat u}}{\bs p}=-\IP{\Delta\bs A\bs{\hat u}}{\bs v},
\end{equation}
which implies that the total weighted population size can be potentially increased if it is possible to find the allowable changes of $\Delta\bs A$, which make the expression \eqref{eq2:4} positive. Hence, at each evolutionary step our goal is to maximize \eqref{eq2:4}. We note that in this case we are not guaranteed to reach the global maximum.

Before we illustrate the process of evolutionary adaptation in this case, we should pause and note two complications that arise here. These complications are also relevant to the application of the described methods to general Lotka--Volterra models.

In the previous section we were able to use the process of evolutionary adaptation because the food chain is globally stable if there exists an internal unique equilibrium. If we allow, as in this section, for the matrix $\bs A$ stops being of the form \eqref{eq1:2}, and assume that $|a_{ij}|>0$ if $|i-j|>2$, then the presence of the internal equilibrium does not guarantee that the system during the fast time scale settles at it; it is actually possible to have much more complicated scenarios in this case, see \cite{hofbauer1998ega}. To make sure that the process of evolutionary adaptation works as desired we must require one more condition at each step: that the system that we obtain after our perturbation is \textit{permanent}.

Recall that general Lotka--Volterra system \eqref{eq1:1} is permanent, if its orbits, after a sufficiently large initial time, end up in a compact set $L\subseteq \Int\R^n_{+}$, where $\Int\R^n_{+}$ is the interior of $\R^n_+$. A necessary condition for \eqref{eq1:1} to be permanent is the existence of unique internal equilibrium $\bs{\hat{u}}\in L$. From the discussion above it follows that for the food chains this condition is also sufficient, but it is no longer true for general $\bs A$. What is most important for us is the fact that if system \eqref{eq1:1} is permanent then
\begin{equation}\label{eq2:5}
    \lim_{T\to\infty}\bs{\bar u}=\lim_{T\to\infty}\frac{1}{T}\int_0^T \bs u(t)\D t=\bs{\hat u},
\end{equation}
i.e., the averages $\bs{\bar u}$ along the orbits are exactly the coordinates of the internal equilibrium in the limit $T\to\infty$. This means that for the permanent systems the coordinates of the unique equilibrium still contain significant information about system behavior even if the system itself is not attracted to $\bs{\hat u}$.

Therefore, for our evolutionary algorithm to work we also at each step should check whether the obtained allowable perturbation implies a permanent food web. There are efficient, but computationally involved, algorithms to perform this task (e.g., \cite{stadler1993probability}), but for simplicity we shall resort to only checking the condition \eqref{eq2:5}, as our numerical experiments show this works well for most of the time. Specifically, we run our system from a number of random initial conditions and calculate time averages along the orbits; if they are sufficiently close to the coordinates of the equilibrium $\bs{\hat u}$ we assume that the system is permanent and we can continue out adaptation.

The second complication is that we consider such changes in the matrix $\bs A$ that non necessarily lead to the true predator--prey relation. Namely, we allow, for instance, $a_{25}$ to be positive and $a_{52}$ be zero, or $a_{52}$ be negative and $a_{25}$ be zero, hence including more general trophic interactions.

Now we are ready to summarize the algorithm for the evolutionary adaptation if matrix $\bs A$ is allowed to be changed. As usual, we start with an allowable matrix $\bs A$, and maximize $\Delta S_p$ in \eqref{eq2:4} assuming that $\Delta a_{ij}$ satisfy \eqref{eq2:1} and \eqref{eq2:2}. Moreover, we also assume additionally that elements of $\bs A+\Delta\bs A$ are such that all the entries above the main diagonal are non-positive, and all the entries below the main diagonal are non-negative. Finally, we check that the new coordinates $\bs{\hat u}+\Delta\bs{\hat u}$ are at distance at least $\delta$ units from the boundary of $\R^n_+$ and that the resulting system is permanent. If all the conditions are met, we can make the next evolutionary step in maximizing the total weighted population size.

\begin{figure}[!h]
\begin{center}
\includegraphics[width=0.5\textwidth]{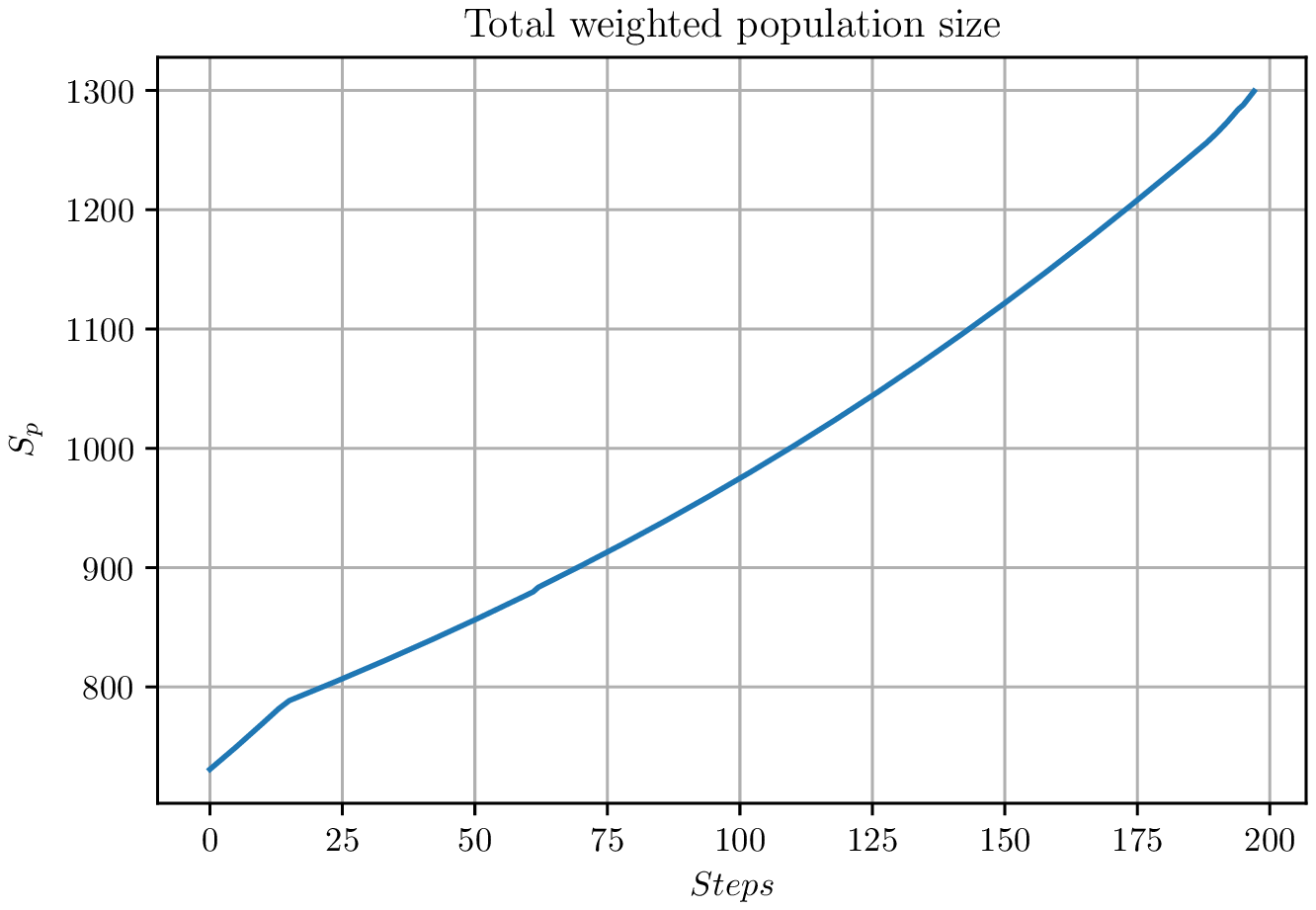}
\end{center}
\caption{The change in the value of functional $S_p$ given by \eqref{eq1:8} at each step of evolutionary algorithm in Section 2. }\label{fig:3}
\end{figure}

To see what may happen in this scenario, we take $\bs A$, $\bs \rho$, $\bs p$ as in \eqref{eq:A}, \eqref{eq:B}, and \eqref{eq:C}. As before we take $\delta=0.001$ and $\varepsilon=0.01$. Following the steps outlined above, we find that our evolutionary algorithm is capable to increase $S_p$ approximately 1.8 times, the final value is $S_p(end)=1300$ (recall that $S_p(0)=731$), Fig. \ref{fig:3}.

\begin{figure}[!h]
\begin{center}
\includegraphics[width=0.5\textwidth]{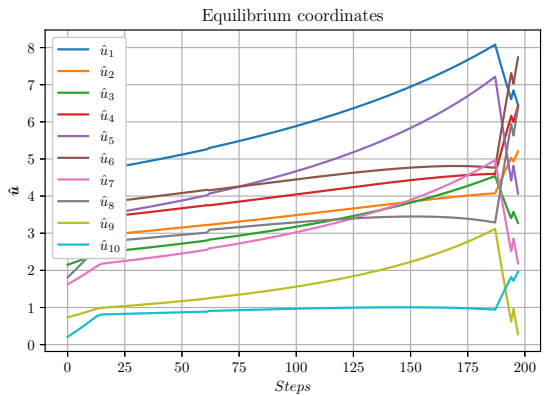}
\end{center}
\caption{The changes in the equilibrium coordinates in the example of evolutionary algorithm in Section 2.}\label{fig:4}
\end{figure}

This time however, the algorithm stops because one of the equilibrium coordinates approaches zero, see Fig. \ref{fig:4}.

\begin{figure}[!h]
\begin{center}
\includegraphics[width=0.5\textwidth]{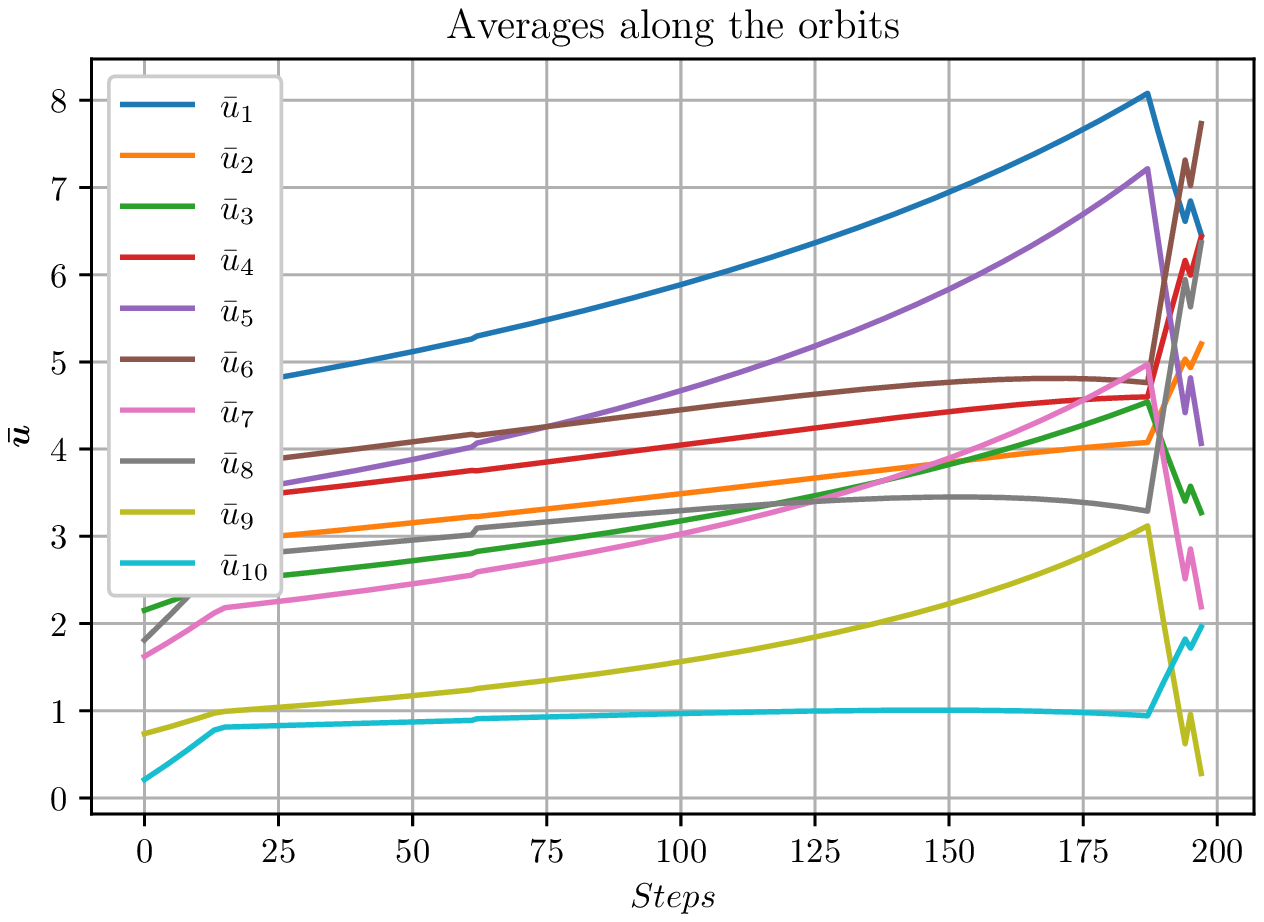}
\end{center}
\caption{The averages along the orbits at each evolutionary step of the evolutionary algorithm in Section 2. Compare with Fig. \ref{fig:4}.}\label{fig:5}
\end{figure}

For comparison we also show the averages along the orbits in this particular case, Fig. \ref{fig:5}.

\begin{figure}[!h]
\begin{center}
\includegraphics[width=0.5\textwidth]{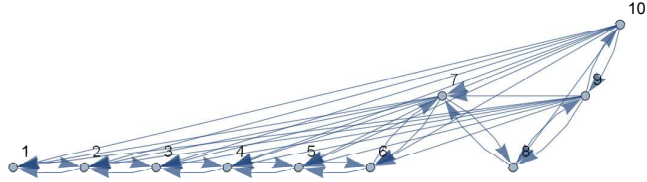}
\end{center}
\caption{The trophic graph of the resulting food web in the numerical example in Section 2.}\label{fig:6}
\end{figure}

Finally, the resulting graph is shown in Fig. \ref{fig:6}, where the numbering of the vertexes corresponds to the linear order of the species in the original food chain, number 12 being the basal species, and number 10 being the top predator. We have both arrows connecting $i$ and $j$ is they are in the relation ``predator--prey'' and only one arrow if the relation of amensalism or commensalism.

\section{Maximizing the mean population fitness in the Lotka--Volterra systems}
In the previous two sections we considered applications of the principle of evolutionary adaptation to the food webs using as an optimization criterion the total (weighted) population size. Such an approach is very natural from the point of view of artificial management of the resources because it is the total population sizes of some of the species in a given food web that are frequently the most important quantities from the point of view of harvesting or survival of some species. From the point of view of natural evolution this criterion is doubtful at best. Hence we would like to consider a different extreme principle for the possible evolution of food webs.

As in many other sciences the search for the extreme principle that shapes the population communities has a long history, see some discussion in \cite{svirezhev1978stability}; for our purposes we \textit{define} the population fitness in the Lotka--Volterra models \eqref{eq1:1} as
\begin{equation}\label{eq3:1}
F(\bs u)=\sum_{i=1}^n \rho_i u_i-\frac 12\IP{\bs{Au}}{\bs u},
\end{equation}
for the given population vector $\bs u$. To the best of our knowledge this function was used to find the correlations of system's dynamics with some extremal principle (namely, maximizing \eqref{eq3:1}) for the first time in \cite{svirezhev1978stability}. As some justification we note that if $\bs A$ is symmetric, then $F$ is increasing along the orbits of \eqref{eq1:1}. Indeed,
\begin{align*}
\dot F(\bs u)&=\sum_{i=1}^n \rho_i^2 u_i-2\sum_{i=1}^n (\bs{Au})_i\rho_i u_i+\sum_{i=1}^n(\bs{Au})_i^2=\\
&=\sum_{i=1}^n(\rho_i-(\bs{Au})_i)^2u_i\geq 0.
\end{align*}
This property is no longer true if $\bs A$ is not symmetric, but we still consider \eqref{eq3:1} as the optimality criterion that we expect to be maximized during the long time evolution of ecological community. Certainly, it is well known that for many mathematical models of evolution the mean population fitness is actually not increasing and not monotone (see, e.g., \cite{birch2015natural,bratus2017adaptive,Bratus2021} for a detailed discussion), but as a first approximation of the action of natural selection on the ecological communities it is arguably the most logical extremal principle.

To summarize, in what follows we apply the principle of evolutionary adaptation to the permanent Lotka--Volterra model \eqref{eq1:1} under the goal to maximize the functional \eqref{eq3:1} at the internal equilibrium $\bs{\hat u}$, allowing now for both the vector $\bs \rho$ and matrix $\bs A$ to be changed during the slow time evolution. The only requirement, as before, is the constraint to keep them in the form to describe food webs.

As before we consider the linear system \eqref{eq1:7}, where now we assume that both the elements of $\bs A$ and $\bs \rho$ depend smoothly on the slow time $\tau$, and also satisfy the constraints
\begin{equation}\label{eq3:2}
    -m_i\geq \rho_i\geq -M_i,\quad i=2,\ldots,n, \quad \sum_{i,j=1}^n a_{ij}^2\leq K,
\end{equation}
for given positive constants $m_i, M_i,K$.

Denote $\Delta\bs A,\Delta\bs{\hat u},\Delta\bs{\rho}$ the main parts of perturbations at each evolutionary step of $\bs A,\bs{\hat u,\bs{\rho}}$ respectively (we omit the dependence on $\tau$). Our first goal is to find the perturbation the fitness functional \eqref{eq3:1} gets in this case.

From \eqref{eq1:7} it follows that
\begin{equation}\label{eq3:3}
    \Delta \bs A\bs{\hat u}+\bs A\Delta\bs{\hat u}=\Delta\bs{\rho}.
\end{equation}
Now,
$$
\Delta F=\IP{\bs\rho}{\Delta\bs{\hat u}}+\IP{\Delta\bs\rho}{\bs{\hat u}}-\frac 12\IP{(\bs A+\bs A^\top)\bs{\hat u}}{\Delta\bs{\hat u}}-\frac 12\IP{\Delta\bs A\bs{\hat u}}{\bs{\hat u}}.
$$
From \eqref{eq3:3} we have
$$
\Delta\bs{\hat u}=\bs A^{-1}(\Delta\bs \rho-\Delta\bs A\bs{\hat u}),
$$
therefore,
$$
\Delta F=\frac 12\IP{\bs \rho}{\Delta\bs{\hat u}}+\frac 12\IP{\Delta\bs \rho}{\bs{\hat u}}=\frac 12\left(\IP{(\bs A^{-1})^\top\bs\rho}{\Delta\bs\rho-\Delta\bs A\bs{\hat u}}+\IP{\bs{\hat u}}{\Delta\bs \rho}\right).
$$
If we introduce an auxiliary linear system
\begin{equation}\label{eq3:4}
    \bs A^\top\bs v=\bs\rho\implies \bs v=(\bs A^\top)^{-1}\bs\rho,
\end{equation}
we finally find
\begin{equation}\label{eq3:5}
    \Delta F=\frac 12\left(\IP{\bs{\hat u}+\bs v}{\Delta\bs \rho}-\IP{\bs v}{\Delta\bs A\bs{\hat u}}\right),
\end{equation}
where the perturbations $\Delta\bs \rho$ and $\Delta\bs A$ must be allowable, i.e., satisfy the conditions \eqref{eq3:2}, keep the structure of the system as for a food web, maintain the uniqueness of the internal equilibrium $\bs{\hat u}+\Delta\bs{\hat u}$, and guarantee that the resulting system is permanent, similar to what we discussed in more details in the previous section.

\begin{figure}[!h]
\begin{center}
\includegraphics[width=0.5\textwidth]{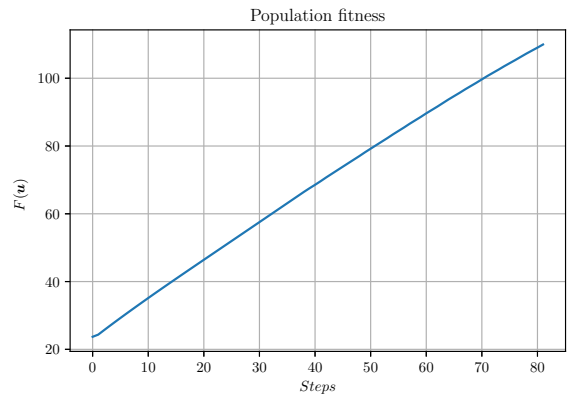}
\end{center}
\caption{Increase of the mean population fitness in the numerical example of evolutionary adaptation to maximize \eqref{eq3:1}.}\label{fig:7}
\end{figure}

To illustrate the algorithm consider the same initial matrix $\bs A$ and initial vector $\bs \rho$ as in the example of Section 1, i.e., \eqref{eq:A} and \eqref{eq:B}. Solving the linear programming problem on each step for maximizing $\Delta F$, we find that it is possible to increase the mean population fitness almost 4.4 times, see Fig. \ref{fig:7}.

\begin{figure}[!h]
\begin{center}
\includegraphics[width=0.5\textwidth]{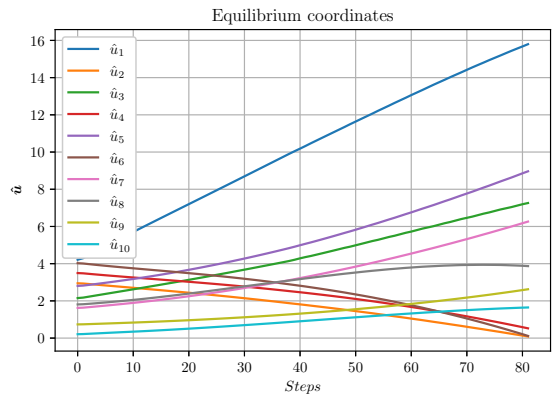}
\end{center}
\caption{The changes in the equilibrium coordinates in the numerical example of evolutionary adaptation to maximize \eqref{eq3:1}.}\label{fig:8}
\end{figure}

For this specific calculation it turns out that at some step the system stops being permanent, which we check in the same way as was described in Section 2. The changes in the coordinates of the equilibrium $\bs{\hat u}$ are shown in Fig. \ref{fig:8}.

\begin{figure}[!h]
\begin{center}
\includegraphics[width=0.5\textwidth]{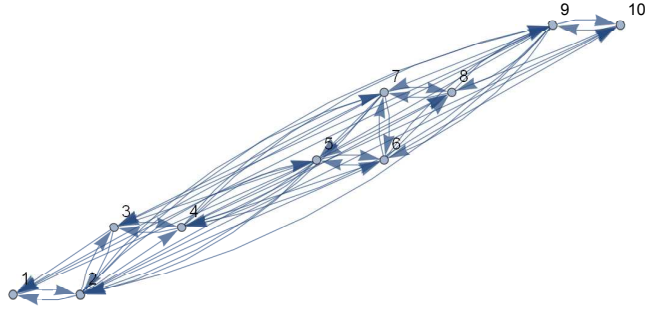}
\end{center}
\caption{The trophic graph of the resulting food web in the example of Section 3. See details in the text.}\label{fig:9}
\end{figure}

As before, we find that the major changes come as a result of turning the initial food chain into a much more complicated structure of a food web, see Fig. \ref{fig:9}. It is an interesting open question to find some regularities in the final trophic web, obtained as a result of evolution under the principle of maximizing the mean population fitness \eqref{eq3:1}.

\section{Concluding words}
Here we presented a relatively simple computational algorithm to incorporate an explicit time dependent community evolution in the classical Lotka--Volterra models. This is an approximate algorithm, however it consists in a sequence of steps at each of which a linear programming problem is solved, which can be done very fast and accurately. We illustrate this algorithm by using two different optimization criteria, namely, optimizing the total weighted  population size and what we call ecological fitness, which can be considered as an analogue of the potential energy in physics. We show that our algorithm performs well being able to increase the required functional in a significant way.

First we apply our algorithm to the food chains and show that it is always possible, by managing the death rates of the system, to increase the total population size of the community. Moreover, a simple algebraic test indicated which death rates should be increased, and which --- decreased.

We also consider more complicated cases allowing for the appearance and disappearance of new trophic interactions. An interesting observation here is that the food chain, which is our usual initial step, is not evolutionary stable under given extremal principals, and it always evolves into a more complicated structure of food web.

It is an interesting open problem to identify the regularities in the obtained under evolutionary adaptations systems, which is part of the ongoing work.

\paragraph{Acknowledgements:} ASB is supported by the Russian Science Foundation Grant 19-11-00008 and
by the Ministry of Science and Higher Education Grant 075-15-2019-1621.

\bibliography{prebiotic}

\end{document}